# A Proposed General Method for Parameter Estimation of Noise Corrupted Oscillator Systems

Francis J. O'Brien, Jr., Nathan Johnnie, Susan Maloney, and Aimee Ross
*Undersea Warfare Combat Systems Department*
*Naval Undersea Warfare Center, Division Newport*
Newport, RI

**ABSTRACT**

This paper provides a proposed means to estimate parameters of noise corrupted oscillator systems. An application for a submarine combat control systems (CCS) rack is described as exemplary of the method.

**Damped Oscillation**

The differential equation modeling damped harmonic oscillators in terms of time *t* is given as the linear second order homogeneous system, derived from Hooke's Law and Newton's Second Law

$$m\frac{d^2x}{dt^2} + c\frac{dx}{dt} + kx = 0$$

If the mass *m* is retarded by a frictional force, $c\frac{dx}{dt}$, proportional to velocity with constant *c*, then dividing by *m*, provides

$$x'' + 2bx' + \omega^2 x = 0, \ (b \geq 0, \omega > 0)$$

where, $2b = \frac{c}{m}$ for mass *m* and arbitrary constant *c*, $\omega = \sqrt{\frac{k}{m}}$.

The characteristic equation is:

$$r^2 + 2br + \omega^2 = 0$$

and the roots are:

$$-b \pm \sqrt{b^2 - \omega^2}$$

In the absence of friction ($b$ = 0) Simple Harmonic Motion (SHM) results.

Damping of the oscillation occurs in one of three ways depending on the value of the discriminant, $(b^2 - \omega^2)$:

- Critical damping $(b = \omega)$→real and equal roots
- Overcritical damping $(b > \omega)$→real and unequal roots
- Undercritical damping $(0 < b < \omega)$ for $\alpha^2 = \omega^2 - b^2$→imaginary roots,
  $r_1 = -b + \alpha i$ ; $r_2 = -b - \alpha i$

**Background**

The focus of this research is modeling undercritical damping or underdamping $(0 < b < \omega)$ in the presence of noise. This is the least desirable situation, for example, in machine design due to the possibly long settling down window of the damping causing excessive vibrations in machine performance.

In this case the real-variable general solution for 2nd order homogeneous DEs with complex conjugates is well known:

$$x(t) = e^{-bt}(C_1 \cos \alpha t + C_2 \sin \alpha t),$$

or redefining the arbitrary constants (Finney/Thomas, Ch. 16),

setting $C_1 = C\sin\varphi$ and $C_2 = C\cos\varphi$ and using identity $\sin(A + B)$:

$$x(t) = Ce^{-bt}\sin(\alpha t + \varphi)$$

where φ is the initial phase angle of the forward motion;

$$\varphi = \text{Arctan}\left[\frac{C_1}{C_2}\right] = \text{Arctan}\left[\frac{C\sin(\varphi)}{C\cos(\varphi)}\right], \begin{pmatrix} -\pi \le \varphi < \pi \\ C\cos(\varphi) \ne 0 \end{pmatrix}$$

calculated at the origin, $t = 0$.

The period of the oscillation is

$$T = \frac{2\pi}{\alpha} = \frac{2\pi}{\sqrt{\omega^2 - b^2}},$$

where α is the damped angular or circular frequency. The period or “pseudo period” crosses the *t* axis twice.

The solution for $T$ is similar to undamped oscillation (simple harmonic motion), ($b$ = 0), except the amplitude is not constant.

Lastly, for reference, we state the derivative and integral of the general solution,
$x(t) = Ce^{-bt}\sin(\alpha t + \varphi)$:

$$x'(t) = Ce^{-bt}[\alpha\cos(\alpha t + \varphi) - b\sin(\alpha t + \varphi)]$$
$$x'(0) = C[\alpha\cos(\varphi) - b\sin(\varphi)]$$
$$x'(t) = 0 \text{ for } t = \frac{\arctan\left(\alpha/b\right) - \varphi}{\alpha} \quad (\text{max, period } 1)$$
$$x''(t) = Ce^{-bt}\left[\left(b^2 - a^2\right)\sin\left(\alpha t + \varphi\right) - 2\alpha b\cos\left(\alpha t + \varphi\right)\right]$$

The indefinite integral is:

$$\int x(t)dt = Ce^{-bt}\frac{1}{\alpha^2 + b^2}\left[b\sin(\alpha t + \varphi) - \alpha\cos(\alpha t + \varphi)\right]$$

As an example, if the noise-free input DE is,

$$x'' + 2x' + (\pi^2 + 1)x = 0$$

and, the initial conditions are,

$$x_0 = x(0) = \sqrt{2} = C\sin(\varphi)$$
$$x_0' = x'(0) = (\pi - 1)\sqrt{2},$$

then the solution of the DE is calculated to be $(C = 2, b = 1, \alpha = \pi, \varphi = \pi/4)$:

$$x(t) = 2e^{-t}\sin\left(\pi t + \frac{\pi}{4}\right)$$

with amplitude-tangent envelope, $e(t) = \pm 2e^{-t}$.

The period is $T = \frac{2\pi}{\alpha} = 2$, obtained in the interval

$$\left[t_1 = t_{\frac{\pi}{2}} = \frac{1}{4}; t_2 = t_{\frac{5\pi}{2}} = t_1 + T = \frac{9}{4}\right]$$

where $x(t), e(t)$ intersect, and $t_{k\pi} = \frac{k\pi - \varphi}{\alpha}$ (see Appendix).

The shift factor or time delay is,

$$\Delta t = \frac{\varphi}{\alpha} = \frac{1}{4}.$$

Note that $x(t) = \pm e(t)$ if and only if $\sin(\alpha t + \varphi) = \pm 1$ which occurs only when $\alpha t + \varphi = \frac{\pi}{2}, \frac{3}{2}\pi,$ etc., a well property of the sine wave. This point-of-intersection property will be exploited in proposed method described below.

Thus, unlike SHM, $x(t) = \pm e(t)$ cannot be determined for the first tangent by finding the first period max., $x'(t) = 0$ which gives $t_{\max} = 0.15$ vice correct value $t_{\frac{\pi}{2}} = 0.25$. This is a common mistake. They are equal only when $b = 0$ since

$$\lim_{b \to 0} \left[ \frac{\arctan\left(\frac{\alpha}{b}\right) - \varphi}{\alpha} \right] = \frac{\frac{\pi}{2} - \varphi}{\omega}$$

which is the max. in the case for SHM. Continuing for all peaks/valleys provides $\pm 1$ values by relation, $\sin x = \cos(x - \pi/2) = \cos(k\pi)$ for $x = \frac{2k+1}{2}\pi$ ( $k = 0,1,2, \ldots$ ).

The following illustrates the underdamping solution with *t* on the *x*-axis:

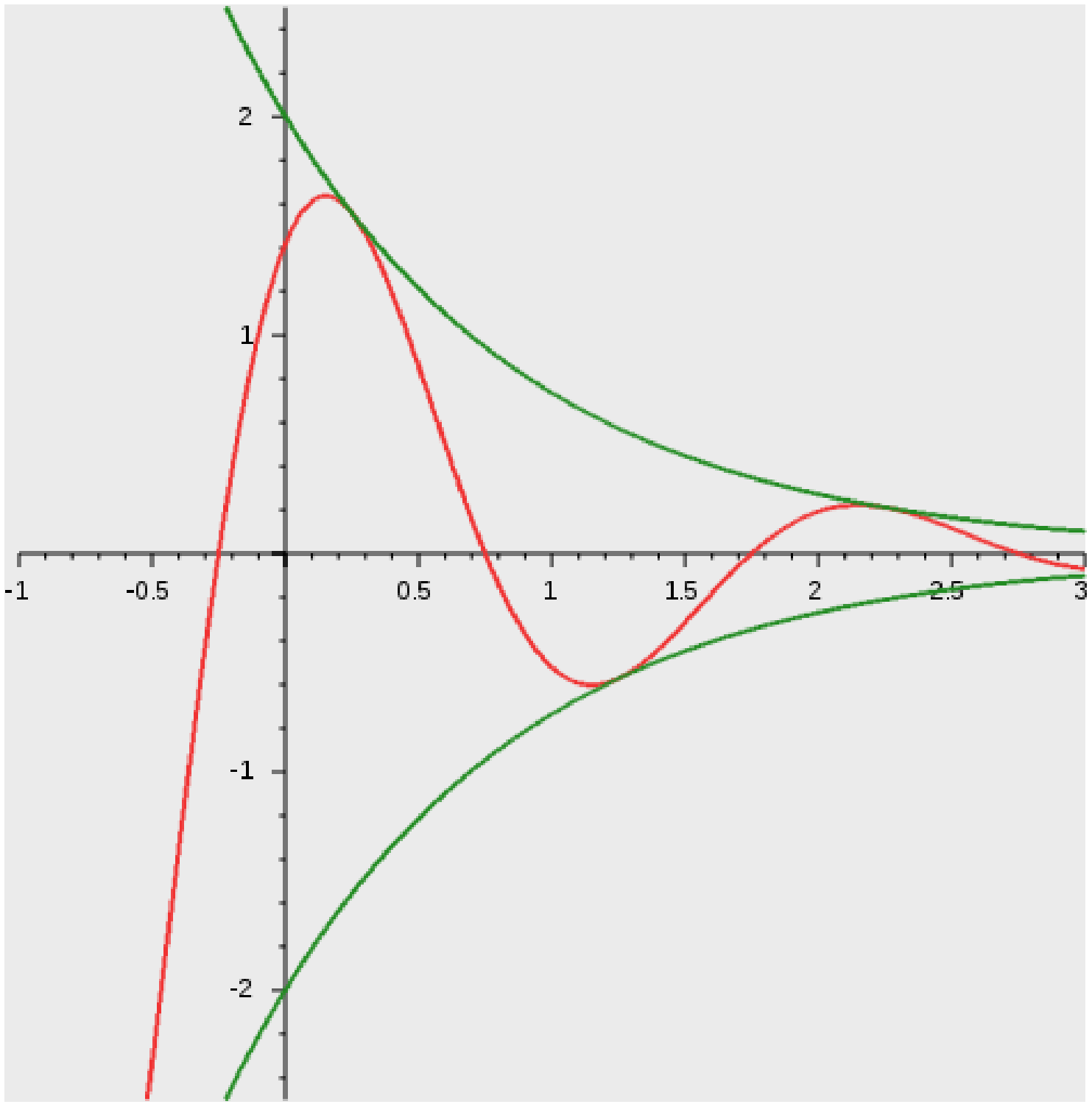

Figure 1. Illustrative solution for $x'' + 2x' + (\pi^2 + 1)x = 0$
NOTE: This solution is used to exemplify calculations in the proposed method.

## PROBLEM STATEMENT

The investigator is initially confronting only a graphical solution such as in Fig. 1 (without the envelope), assumed as abstracted from filtered noisy input data. The requirement is to estimate the parameter set $[T, C, \alpha, b, \varphi]$ to characterize the noise-corrupted oscillatory motion in mechanical and electromagnetic systems. That is, we assume no known DE or initial conditions as inputs governing the observed oscillations corrupted by unknown noise type and degree. This is not an easy problem that is solvable mathematically in closed form. It is a challenging engineering problem to solve. For example, common sources of mechanical vibrations include multiple factors:

- Time-varying mechanical force/pressure
- Fluid induced vibration such as intermittent wind, tidal waves, etc.
- Acoustic and ultrasonic
- Random movements of supports such as seismic
- Thermal, magnetic, etc.

Fourier series analysis only represents true periodic functions,

$$f(x+p) = f(x), \text{period } p$$

so this tool is not explicitly available for noise corrupted non–periodic damped oscillation parameter estimation (Boas, Ch. 7, Fourier series and transforms). A number of analysis protocols have been proposed for non–periodic analysis including nonlinear estimation, discrete Fourier-Series Transformation (DFT) of signals sampled asynchronously, and others. However, many digital signal processing algorithms used for parameter estimation and digital measurement technology have been used often with unsatisfactory accuracy due to a number of factors (noisy data, digitalization, unsuitable sampling conditions, sensitivity to initial conditions, approximate mathematical methods, and other factors).

A proposed estimation method is described for this common problem important in the areas of shock/vibration analysis, electrical circuits, as well as other areas of applications in military science and engineering, and commercial applications. A specific naval application for submarine combat systems is given below. The proposed technique comes from the review of simulation studies conducted at NUWC Newport in the last several years.

The standard "formal" method is cumbersome and based on restrictive assumptions that may not be reasonable for real world noisy data; for example, see the online paper from Technologies GmbH, "Precise Parameter Determination of Damped Oscillation Signals".

## Preliminary Data Smoothing

The initial noisy time series data set must be reduced to a manageable mathematical function in order to model the data parametrically. To accomplish this, the time series must first be smoothed by means of a data reduction filter.

Many choices exist for the analyst including:

- Moving Window Average
- Savitski-Golay Moving Average (polynomial least squares)
- FFT low pass filter
  - This filter allows cutting the high frequencies of a signal.
- FFT high pass filter
  - This filter allows cutting the low frequencies of a signal.
- FFT band pass filter
  - This filter allows cutting the low and high frequencies of a signal.
- FFT block band filter
  - This filter allows keeping the low and high frequencies of a signal

To model a damped sinusoidal 2$^{nd}$ order differential equation with small to moderate additive noise in the proposed method, the authors smoothed the data by use of a moving average (MA) which is a type of discrete-time finite impulse response filter (FIRF). We believe the MA is an often overlooked first smoothing process tool which provides meaningful insight into data structure for parameter set, $[T, C, \alpha, b, \varphi]$. Citing a common textbook:

> In spite of its simplicity, the moving average filter is *optimal* for a common task: reducing random noise while retaining a sharp step response. This makes it the premier filter for time domain encoded signals. …. Many scientists and engineers feel guilty about using the moving average filter. Because it is so very simple, the moving average filter is often the first thing tried when faced with a problem. Even if the problem is completely solved, there is still the feeling that something more should be done. This situation is truly ironic. Not only is the moving average filter very good for many applications, it is *optimal* for a common problem, reducing random white noise while keeping the sharpest step response. (Smith, Chapter 15, "Moving Average Filters")

The level of smoothing is indicated by the notation MA-*k* where *k* represents the averaging level. The MA is commonly used with time series data to smooth out short–term fluctuations and highlight longer–term trends or cycles/periodicities. It is also similar to the low–pass filter.

The authors are unable to prescribe the optimal MA smoothing parameter, as this is an unsolved issue in data processing. The most practical engineering solution is to run various levels of smoothing and choose the smoothing MA parameter *k* which provides the smallest root-mean square (RMS) value between the data and the filter, measured vertically at each time–point. That is, compute the RMS measure for $x_j$ for observations *j*=*k* to *N*,

$$RMS(k) = \sqrt{\frac{\sum_{j=k}^{N}\left(x_j^m - x_j\right)^2}{N-k}},$$

where $x_j^m$ is the $j^{th}$ MA-*k* averaged value of *N* observations. Observations that are less than or prior to the values of $x_j^m$ are ignored since they are zeroed out in the MA filter . In comparing several smoothing parameters, select min $[RMS(k)]$ to represent the best fitting smoothing model to the observed noisy data.

**NOTE:** The standard MA-*k* filter uses arithmetic averaging for smoothing. In the alternative other measures of central tendency may work better such as the median value of each overlapping segment; this may better smooth high noise data sets. Alternatives are stated below regarding other smoothing options.

**NOTE:** In previous research, the MA method (using $k = 5$ and 10) was shown to be quite accurate for modeling SHM (low frequency) with 50% Gaussian noise; the overall accuracy was 97-99% using 2 periods averaging. (O'Brien and Johnnie, 2011).

**Demonstration of Model Parameter Estimation from the FIRF for Hypothetical Empirical Data**

In the following we propose a general approach that has been found satisfactory for damped and undamped (simple harmonic motion) oscillation. A detailed explanation is derived and described herein. The method is based on analysis and simulation data run in MATLAB. The error of estimation against known solutions is ≤ 5% on average. We show the traditional solutions as well as the new proposed solutions (Appendix).

We illustrate the implementation of the procedure for the above mentioned damped oscillation DE model, $x'' + 2x' + (\pi^2 + 1)x = 0$, with derived solution,

$$x(t) = Ce^{-bt}\sin(\alpha t + \varphi) = 2e^{-t}\sin\left(\pi t + \frac{\pi}{4}\right).$$

The objective is to estimate each of the model parameters $(T, C, b, \alpha, \varphi)$ for a hypothetical data set of measurements from the graphed solution in Fig. 2. We assume no knowledge of the DE, initial conditions, or its solution, or of the envelope (green curve in Fig. 2). That is, we assume the red curve only has been provided by the FIRF as the best fitting MA filtered solution with no other information[1]. That is, the input model is:

$$x(t) + \eta(t)$$

[1] Note: in an MA FIRF plot the first *k* -1 values are ignored. See O'Brien/Johnnie (2011).

where $\eta(t)$ represents additive noise. The envelope function $e(t) = Ce^{-bt}$ is derived from parameter estimates $b$, $C$.

In this example we know ground truth is the solution set:

$$\begin{cases} T = C = 2 \\ b = 1 \\ \alpha = \pi \\ \pi/4 = 0.79\,(\text{rad.}) \end{cases}$$

The estimated results will be compared using the proposed method against the ground truth values to provide an error bound for the estimated parameters. Figure 2 below provides the numerical information used in the derivation of the estimates. In practice, the numerical values are calculated in automated fashion. The estimates below are based on a single period of data, but averaging across multiple periods reduces errors in the approximations in the presence of noise. The Appendix provides the averaging formulas used across multiple periods. A key factor in the accuracy of the approximations is the best fitting MA solution (or other data reduction filters) to the noisy data and the use of appropriate averaging of estimates across multiple periods.

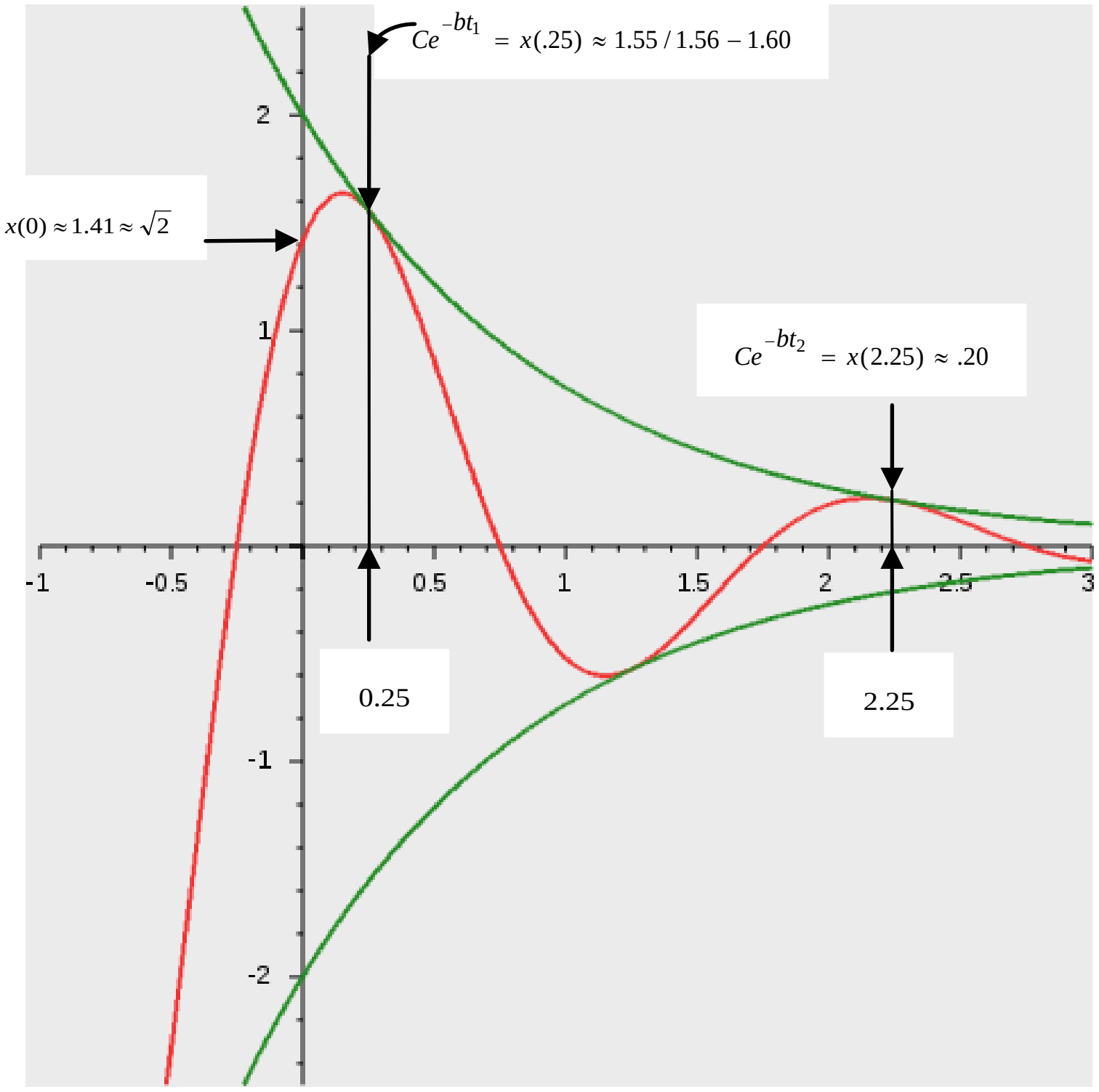

Figure 2. Hypothetical Idealized FIRF Fit for Data from Model $x(t) = 2e^{-t} \sin\left(\pi t + \frac{\pi}{4}\right)$. In actual engineering practice the first *k*-1 values are zeroed out but this does not changes the illustrative calculations in a material manner. $e(t)$ is not available from MA filtering.

## *Model Estimates*

The sequence of calculations of the derived model estimates in the algorithm is shown below. Some calculations are standard (e.g., GmbH paper); some are alternative proposed means. Rounding errors exist due to reporting calculations to two-point accuracy.

The complete algorithm is presented in the Appendix where is provided more accurate estimates with less restrictive assumptions that may not be reasonable for real world noisy data as mentioned earlier. The proposed algorithm proceeds in the reverse (more logical order) compared to the following traditional procedure for obtaining the solution set.

***<u>Damping Factor</u> b:***

This negative quantity in the varying amplitude $Ce^{-bt}$ is typically determined in standard fashion by the following relation with empirical data substituted:

$$b = \frac{1}{t_2 - t_1} \ln\left[\frac{Ce^{-bt_1}}{Ce^{-bt_2}}\right] = \frac{\ln\left(e^{-bt_1}\right) - \left(\ln e^{-bt_2}\right)}{t_2 - t_1}$$

where,

$t_1 = t_{\frac{\pi}{2}}$ ; This is read from Figure 2 as about 1/4

$t_2 = t_{\frac{5}{2}\pi}$ ; This is read from Figure 2 as about 9/4

$Ce^{-bt_1} = x\left(t_{\frac{\pi}{2}}\right)$ is read from the graph to be about 1.55 or 1.56 to1.60 and

$Ce^{-bt_2} = x\left(t_{\frac{5\pi}{2}}\right)$ is about 0.20.

Only $Ce^{-bt_1}$ and $Ce^{-bt_2}$ are assumed known quantities from the smoothing graph, Fig. 2.

From these data, we get

$$b = \frac{1}{2.25 - 0.25} \ln\left(\frac{1.56}{0.20}\right) \approx 1.03$$

or

$$b = \frac{1}{2.25 - 0.25} \ln\left(\frac{1.60}{0.20}\right) \approx 1.04$$

The error of estimation is$\leq 5\%$.

**NOTE**: We question the availability of $Ce^{-bt_1}$ and $Ce^{-bt_2}$ at this point in the modeling since $e(t)$ cannot logically be constructed with only the damping parameter $b$, and the read-off points $t_{\frac{\pi}{2}}$ & $t_{\frac{5}{2}\pi}$ have not been established as explained earlier since both $\varphi$ & $\alpha$ are needed for such a determination. Averaging formulas for more than two points per period improve the accuracy of the estimation for noisy data.

***<u>Constant C</u>:***

Two ways may be used to obtain $C = e(0)$ . First, the constant can be determined from the relation:

$$Ce^{-bt}\Big|_{t=t_1} = x\left(t_{\frac{\pi}{2}}\right)$$

or,

$$C = \frac{x\left(t_{\frac{\pi}{2}}\right)}{e^{-bt}\Big|_{t=t_1}} = \frac{1.56}{e^{-.25}} = 2.00$$

If $x(t)$ is 1.60 at that point, then $C \approx 2.05$. The error of estimate is $\leq 5\%$.

Second, *C* can be determined analytically from the following relation using *b* estimate,

$$C = \frac{1}{2}\left[\frac{Ce^{-bt_1}}{e^{-bt_1}} + \frac{Ce^{-bt_2}}{e^{-bt_2}}\right]$$

with FIRF values substituted. This calculation shows, $C = 2.00 - 2.03$, which is a 3% error depending upon the value used for *b*.

**NOTE***:* We again question the integrity of the logic and show a better technique in the Appendix.

**NOTE**: At this point the envelope function $Ce^{-bt}$ can be written. The envelope function can be drawn by noting that for solved parameters $C = 2, b = 1,$ that $x(t)\,\&\,Ce^{-bt}$ intersect (are tangent) at interval points, $t_{\frac{k}{2}\pi} = t_{\frac{\pi}{2}} + \frac{k-1}{2}$ ($k$ odd) or, $t = \frac{1}{4}, \frac{5}{4}, \frac{9}{4}, \frac{13}{4},$ etc. corresponding to the intersection points of $x(t)$ and $e(t)$ on the sine wave by the angle function theorem for sine. Fig. 2 bears this out.

**Phase Angle $\varphi$:**

Phase angle is measured in one of two ways.

a) From the standard definition on a right triangle with *C* as the hypotenuse, and $x(0) = C\sin(\varphi)$ in Fig. 2,

$$\tan\varphi = \frac{C\sin\varphi}{C\cos\varphi} = \frac{C\sin\varphi}{\left[\sqrt{C^2\left(1-\sin^2\varphi\right)}\right]},$$

or if $C = 2$,

$$\varphi \approx \arctan\left\{\frac{\sqrt{2}}{\left[\sqrt{C^2\left(1-\sin^2\varphi\right)}\right]}\right\} \approx \arctan\left(\frac{\sqrt{2}}{\sqrt{2}}\right) \approx \arctan(1) \Rightarrow \varphi \approx \frac{\pi}{4} = 0.79\ (45^\circ)$$

If estimate $C \approx 2.05,$ we calculate $\varphi \approx 0.76\left(43.6^\circ\right) \approx \frac{6\frac{1}{20}}{25}\pi$ , representing a 3% error.

b) A second method is derived from the measurements $C = 2$ and $x(0) = C\sin(\varphi) \approx \sqrt{2}$ ,

$$\varphi \approx \arcsin\left(\frac{\sqrt{2}}{2}\right) \approx \frac{\pi}{4} = 0.79.$$

If $C$ is 2.05, then $\varphi \approx 0.76,$ as above in a).

**<u>Frequency</u>** $\alpha$

To estimate $\alpha$, we find the point of intersection at $t_1$ between the quantities $x(t) = Ce^{-bt}\sin(\alpha t + \varphi)$ and the envelope function $e(t) = Ce^{-bt}$ , and solve algebraically for $\alpha$.

Equating $x(t) = Ce^{-bt}\sin(\alpha t + \varphi) = e(t)$ or,

$$1 = \sin(\alpha t + \varphi), \text{ and}$$

$$\arcsin(1) = \frac{\pi}{2} = \alpha t + \varphi \Rightarrow \alpha \approx \frac{\pi/2 - \pi/4}{.25} = \pi,$$

using estimate $\varphi = \frac{1}{4}\pi.$

**NOTE**: Solving for $t$ provides the first positive point on the graph (1/4) of the envelope function. Alternatively, it is clear that all $\pm\ e(t)$ points can be computed by calculating,

$$t_{\frac{(2k+1)\pi}{2}} = \frac{t_{\pi k} + t_{(k+1)\pi}}{2} = \frac{(2k+1)\pi - 2\varphi}{2\alpha}$$

which equals $k+\frac{1}{4}(k=0,1,2,\ldots)$ for our data set.

This function represents the average time-point value of a half-period that gives the envelope function $e(t)$, since the sine term drops out of $x(t)$ by noting,

$$\sin\left[\alpha\left(t_{\frac{(2k+1)}{2}\pi}\right)+\varphi\right]=\sin\left[\pi\left(k+\frac{1}{2}\right)\right]=\cos(k\pi)=\pm 1$$

If phase angle is about 0.76, frequency is estimated to about 3.25 ($\pi$ + 0.10) a 3% error from known $\alpha$.

In general, by this approach, if $k$ is an odd integer, then frequency is:

$$\alpha=\frac{k\frac{\pi}{2}-\varphi}{t}\Bigg|\, t=t_{\frac{k}{2}\pi}$$

**NOTE:** If the envelope is not available, set $x(t)$ to a specific value from Fig. 2 and solve by arcsine using derived estimates of $C$, $b$, $\phi$.

An improved method is described in Appendix in terms of time delay.

**<u>Damping Period</u> *T*:**

$$T=\frac{2\pi}{\sqrt{\omega^2-b^2}}=\frac{2\pi}{\alpha}=\frac{2\pi}{\pi}=2 \text{ or about } \frac{2\pi}{(\pi+0.10)}$$

per above estimates.

The estimate is made in the interval; $[t_1,t_2]=\left[\frac{1}{4},\frac{9}{4}\right]$. The upper boundary $9/4$ is selected to be the value of the damped sine wave loop after the second 0-crossover of $t$, $2+\frac{1}{4}$ .

This calculation provides:

1. $\omega^2=\pi^2+1$ or about $(\pi+0.10)^2+(1.01)^2$
2. The reconstructed DE is,
   $x''+2bx'+\omega^2x=0,$
   $x''+2x'+(\pi^2+1)x=0,$
   or about $x''+2.05x'+[(\pi+0.10)^2+(1.01)^2]x=0$ as worst case solutions.
3. Roots are $-b\pm ai=-1\pm i\pi$ or $-1\pm i(\pi+0.10)$
4. Integration will show for the estimated parameters of the period

$$I=\int_{t=0.25}^{t=2.25}\sin(\alpha t+\varphi)dt\approx 0.123.$$

Unlike SHM, $I$ is not zero over $T$.

In summary, the above standard traditional method provides the solution with an error rate $\leq$ 5% or higher. Fig. 3 shows the closeness of the solution between actual and estimated solutions using the worst case estimates. However, as mentioned, we question the logic of the sequence of calculations in the traditional algorithm.

The Appendix describes a general algorithm with less error and based on fewer assumptions and useful averaging formulas across multiple periods. The error rate for the same data was found to be 1–2%.

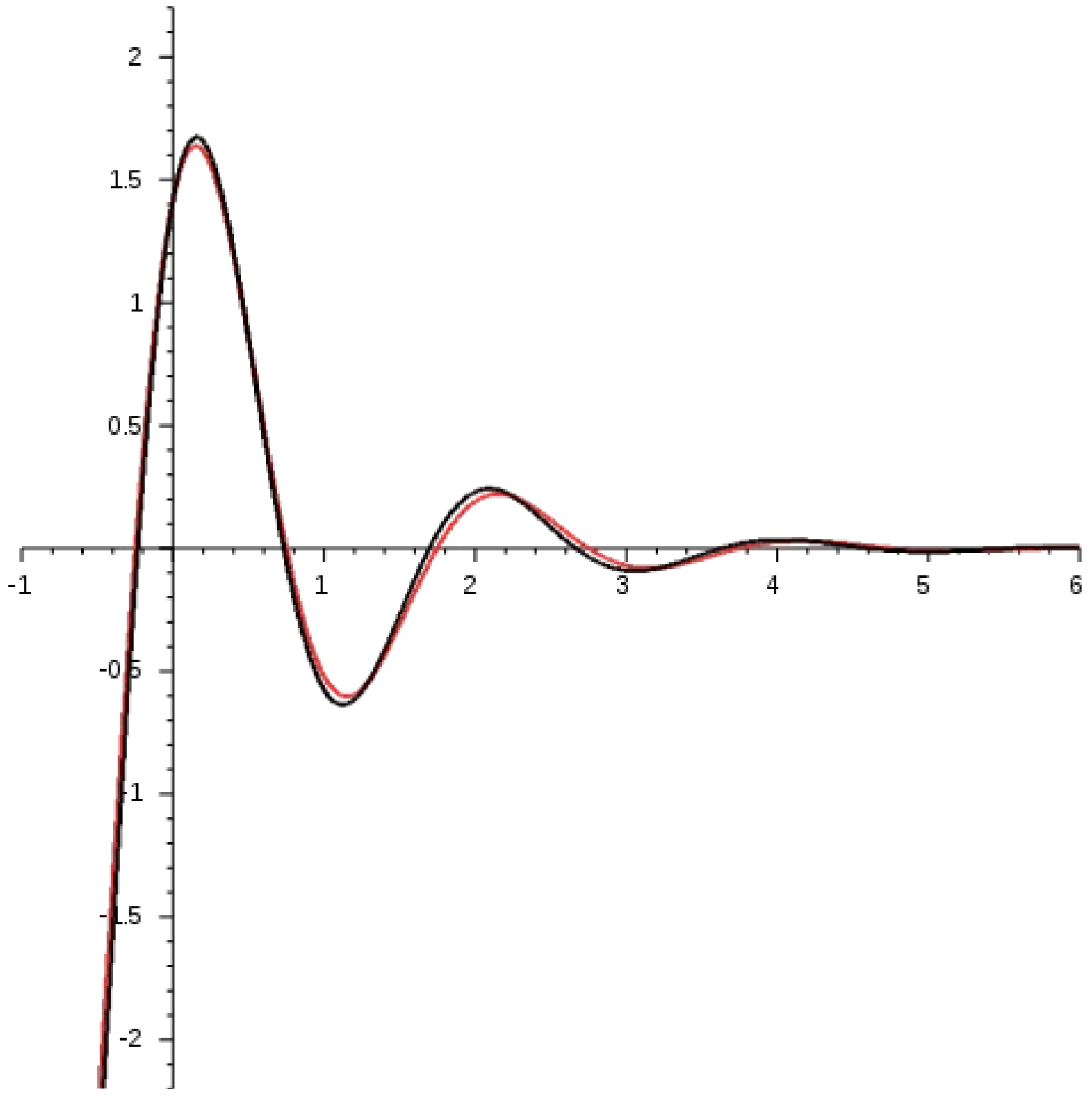

Figure 3. Comparison of True Solution (red) and Estimated Solution (black)

## Application to Navy Engineering Problem

**(UNCLASSIFIED)**

*Engineering Report of a Noisy Combat Systems Submarine Rack*

Case study:

A chassis consists of three individual units A1, A2 and A3 in Figure 4. The main equipment cabinet and the units are supported with individual cooling fans. When the main power is turned on a measureable annoying audible noise is produced by the cooling fans. We can model the disturbance by the DE method of this paper.

Add a 1U or thinner in height–sized rack and mount server to the new generation of equipments. This unit will play an important role in extending the longevity for each chassis, as well as providing technicians with anticipatory data prior to potential problems developing in the system.

The current temperature sensors—labeled “Unsafe Temp” and “Over Temp”—monitor the overall cabinet temperature. This is insufficient to ensure the units operate in normal temperature environment.

In addition to the presence of noise in the system the following could be considered sources of failure:

1. Rise in temperature within the unit.
   - A lock in airflow caused by some restrictions that are not easy to be visualized without some type of sensors.
2. Fluctuations in the relative humidity, which makes the unit more vulnerable to Electrostatic Discharge (ESD).
   - This factor ultimately causes the unit to experience gradual degradation and failure.
3. Condensation could cause resistance, impedance to fluctuate if not controlled properly.
   - This normally is caused by temperature and atmospheric pressure changes.

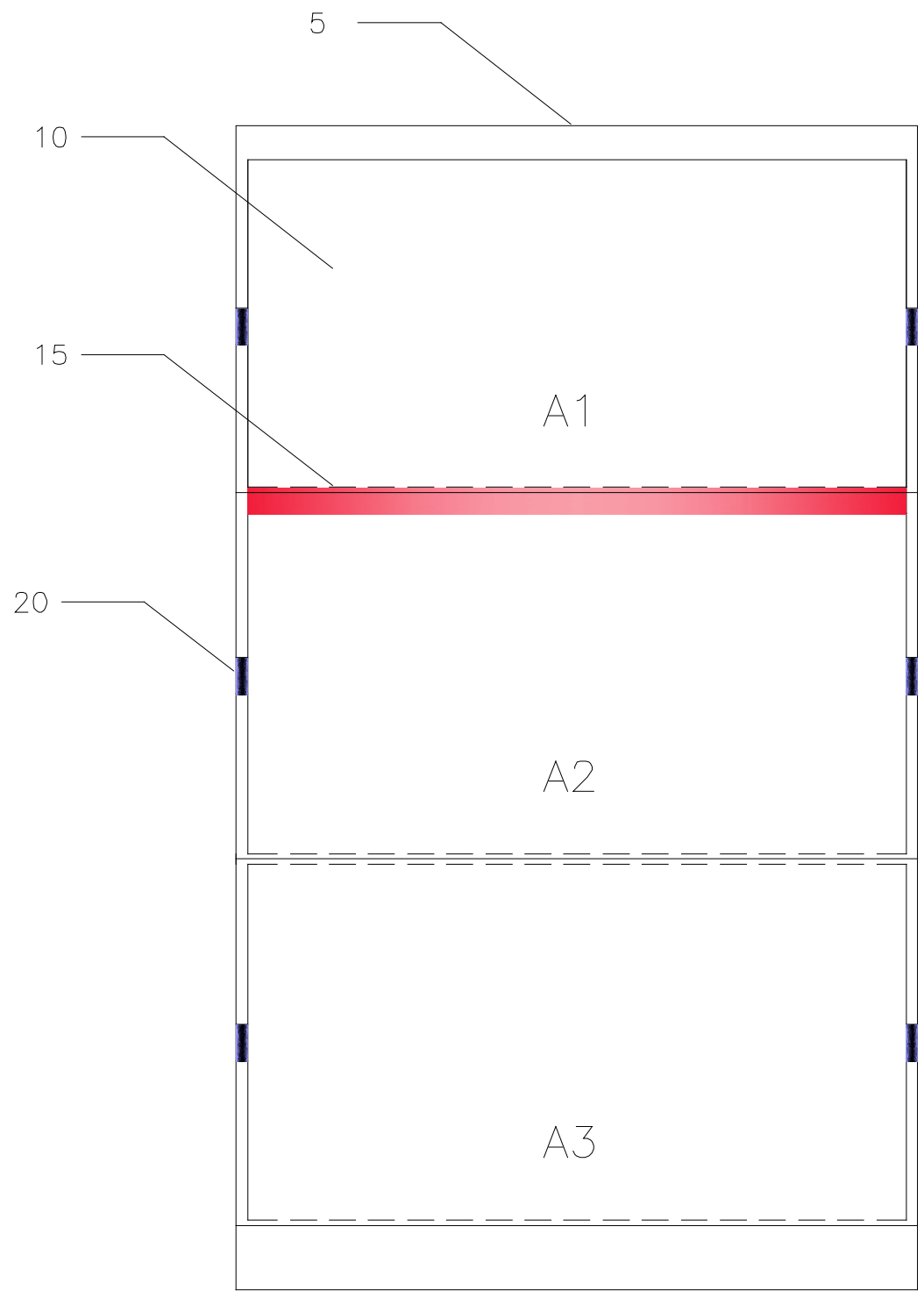

Figure-4 Equipment Cabinet. Note: Damping material is in red.

*Legend*

5 - Enclosure designed to hold several units.

10 - Units A1, A2 and A3 placed inside 5.

15 - B or damping represents dissipation. The cavity between the two units could be a source for unwanted noise. In order for damping to occur, the vacant area is filled with flexible sponge (red region). The noise is indirectly proportional to damping coefficient (*b*).

20 - K is resistance or stiffness of elements that holds any of the units in the enclosure. The stiffness is proportional to displacement. There is a pair of Ks per unit. The unit measurement for stiffness is N/m or N-m/rad.

Investigation:

The noise is noticeable upon powering the chassis with the associated units. The system's main cooling operation is normal and not included in the analysis. The system with disturbance is simplified in the block diagram Figure-5. Measurements indicate that the source of the unwanted noise in the mid unit, cabinet A2. The cooling fans of the unit A2 produce a discrete periodic low frequency audio signal. The noise is continuous and periodic. The sound is similar to drum beats.

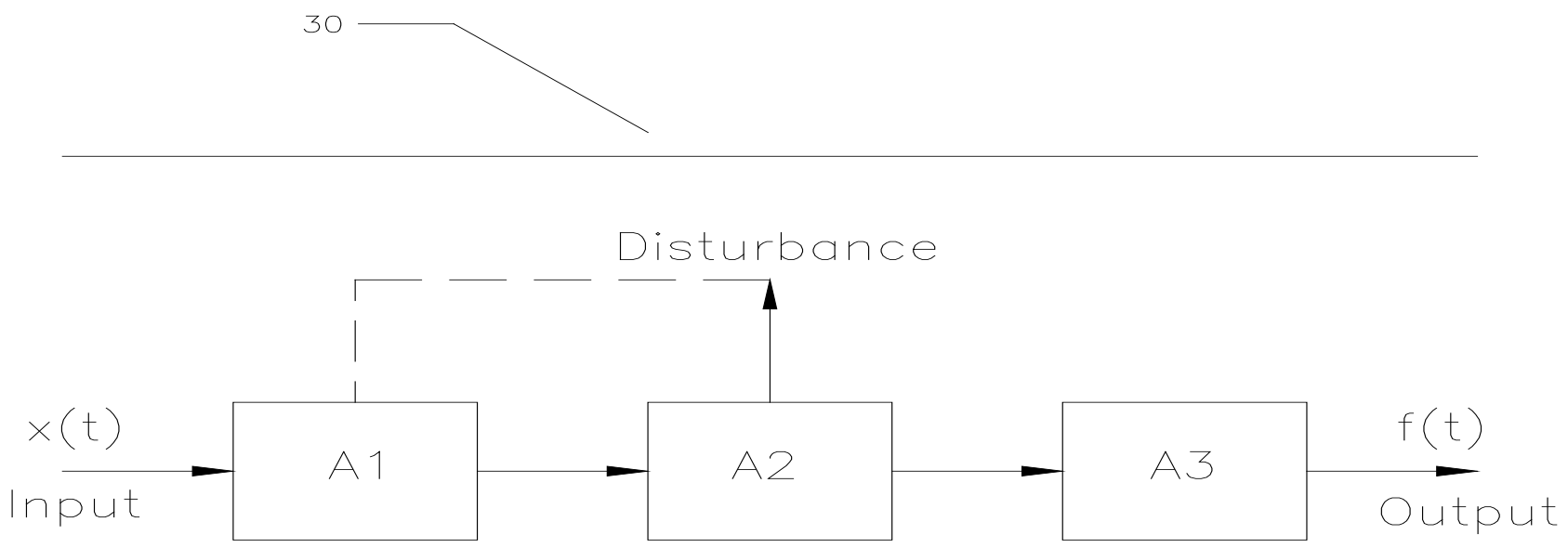

Figure-5 System Block Diagram

*Legend*

30 - System configuration for the units is set in series with disturbance (D) present in unit A2. $x(t)$ is a variable which represents displacement. $f(t)$ represents the applied force on the units.

Cause:

It appeared that the cooling fans in A2 were causing the top cover to vibrate. As a results a low frequency waveform was generated. The waveform collides with the bottom cover of the unit above, A1, causing reverberation. The low frequency audio return called echo inherits a phase shift due to the distance in between units A1 and A2 such as shown in Figure-6.

```
% MATLAB Code for Fig. 6

%
   clear all;
   close all;
   clc;
%  Define the amplitude
   A = 1;
%  sampling frequency
   Fs = 4800;
%  Define the frequency in Hertz
   w = pi;
%  Define phase shift
   theda = .75*pi;
%  Define the time vector
   t = 0:.025:5;
```

```
%   Compute y(t), the sine wave
    signal1 = A * sin(w * t);
%   Compute y(t), the sine wave
    echo = A * sin(w * t + theda);

%   adding the two signals
    Sum_Signal = signal1 + echo;

%   Plot y vs. t
%   subplot(2,1,1);
    plot(t,signal1,'-r*',t,echo,'-.b',t,Sum_Signal,'k');
%   Label x-axis
    xlabel('t (seconds)')
    ylabel('Signal1& Echo ')
    title('Signal1, Echo, Sum Signal = Signal1 + Echo')
    legend('(a) Sig (A1)','(b) Echo (A2)','Sum(a + b)')
```

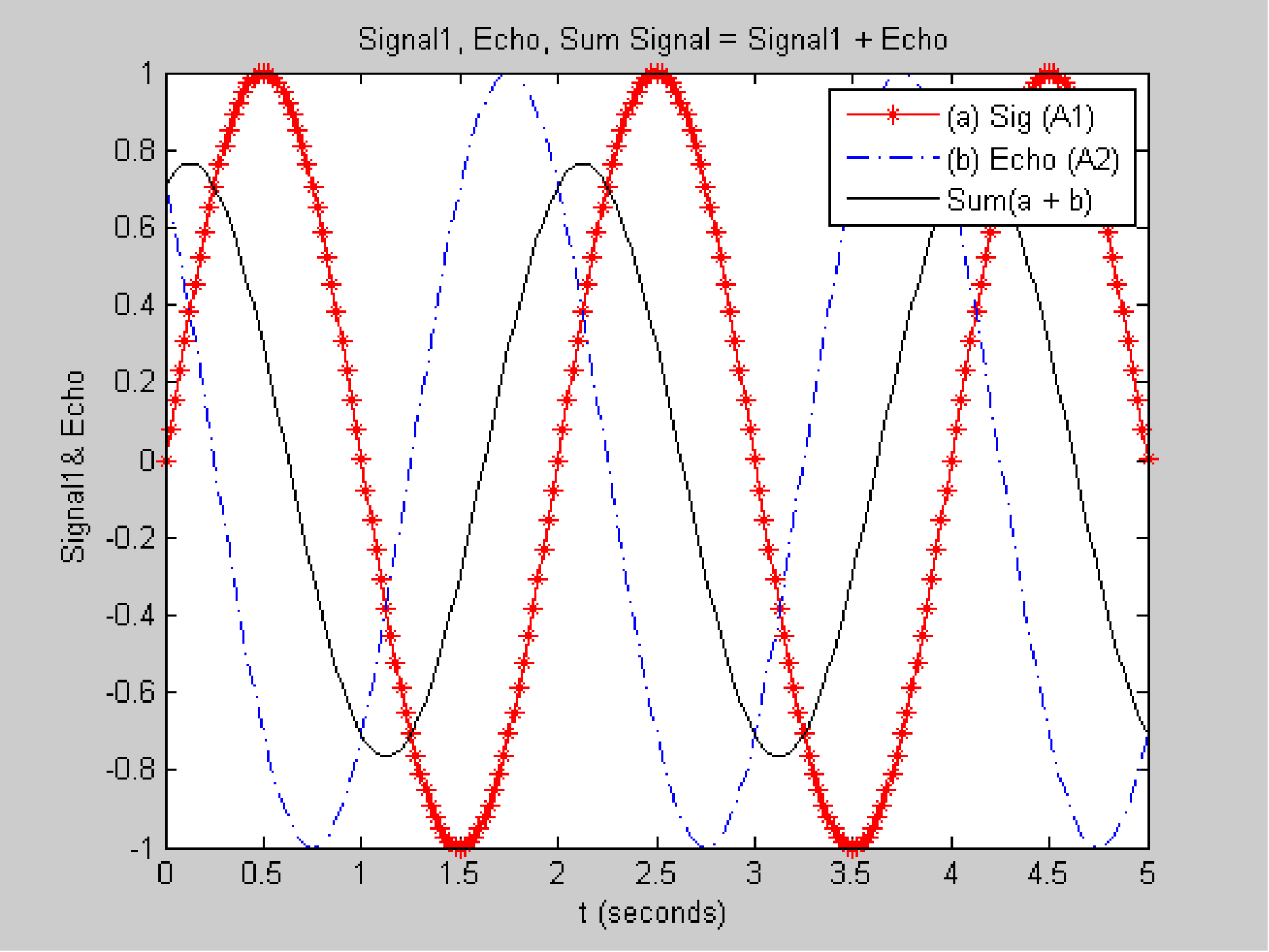

Figure-6 Signal Vs Echo with Summation

*Resolution*:

The solution consisted of placing a sponge type material that fits fully the four corners (top section) of unit A2, Figure-4 and Figure-7. The sponge is resistive and is immune to Electrostatic Discharge (ESD). The thickness is slightly higher than the distance between Unit A1 and the upper unit, A2. The fused establishment impacts absorption of the vibrations caused by the top cover of unit A2. The process is called damping (parameter *b* in the mathematical model). The hardness level of the sponge is both dense and medium soft.

We take advantage of the situation and add several steps to improve the functionality of each chassis:

1. Add temperature sensor to each unit.
2. Add ability to monitor the operation of the fans within each unit.
3. Add ability to monitor continuously the air cooling system, in units $\frac{ft^3}{min.}$ (CFM).
4. Add sensors capable of predicting condensation before it occurs.
   - Especially prior to opening the rack door.

**NOTE 1:** The additions described should be placed where the Power Control Panel is located in such a way that will help the operators or technicians to continuously monitor the rack performance through visual contact in order to observe any developing system abnormalities.

**NOTE 2**: It is important to monitor the humidity outside the cabinet to match the temperature inside the unit to determine the risk of condensation when the cabinet door is opened.

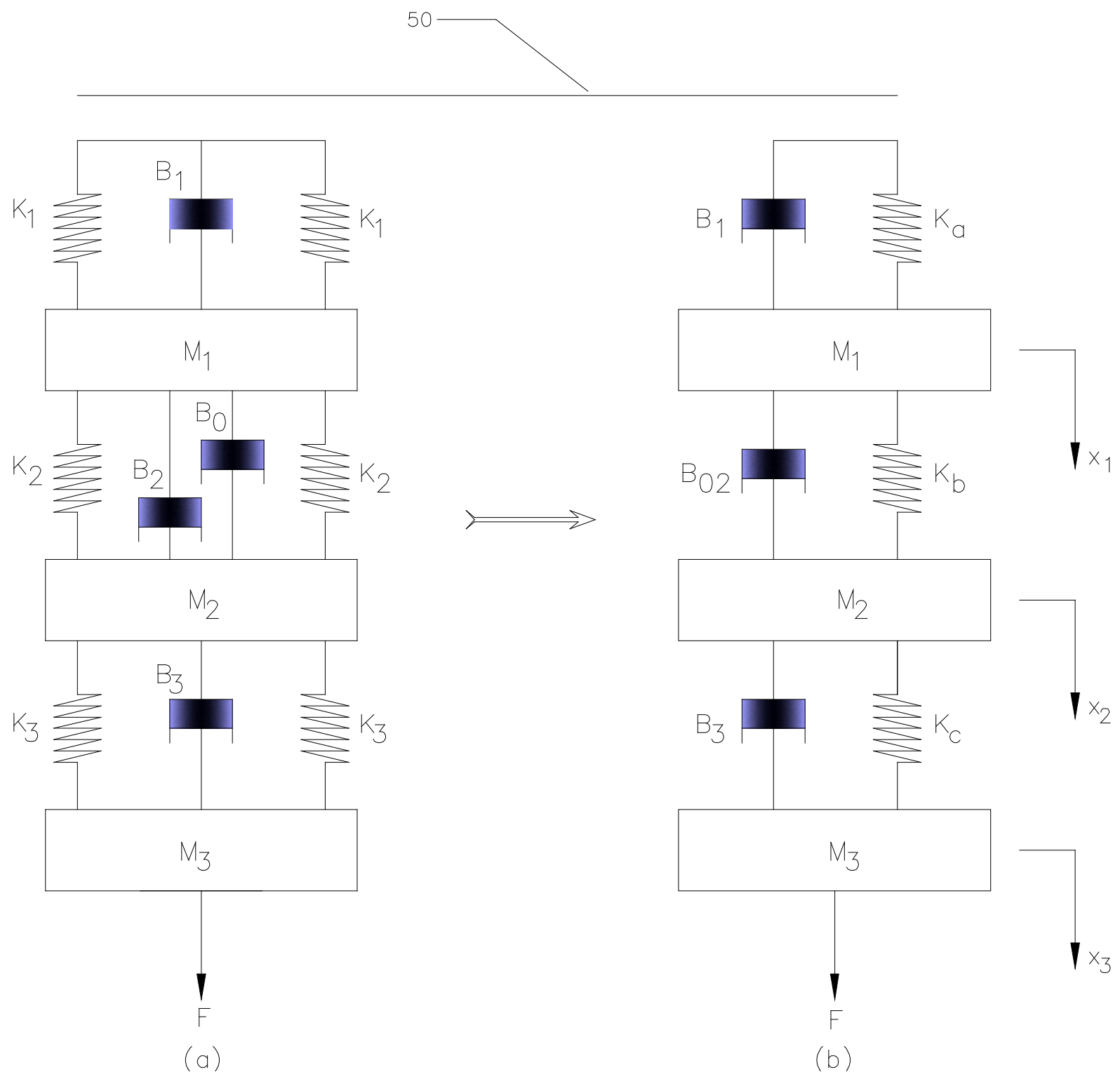

Figure-7 (a) & (b) Mechanical System

*Legend*

50 - Mechanical system representing the inner units of the cabinet alone. It is assumed that the stiffness values for each are equal, or balance equally.
Therefore, since the stiffness in both sides of unit A2 are in parallel, $K_b = K_2 + K_2$
The same process holds true for $K_a$ and $K_c$. The damping coefficients are assumed to be the same in units A1 and A3. Moreover there is natural damping for the units B1, $B_2$ and $B_3$. The $B_2$ damping is not capable of suppressing the disturbance caused by unit A2. $B_0$ is set in parallel with $B_2$ so as to subdue disturbance D as depicted in Figure-5. The $B_{02}$ is calculated to be: $B_{02} = B_2 + B_0$. Thus damping is inversely proportional to the produced audio frequency. In other words, increasing B means lowering frequency. $M_1$, $M_2$ and $M_3$ represent the mass of units A1, A2 and A3 respectively.

Electrical Modeling:

The analogous electrical system for the units is depicted in Figure-8. For simplicity only the three units are considered.

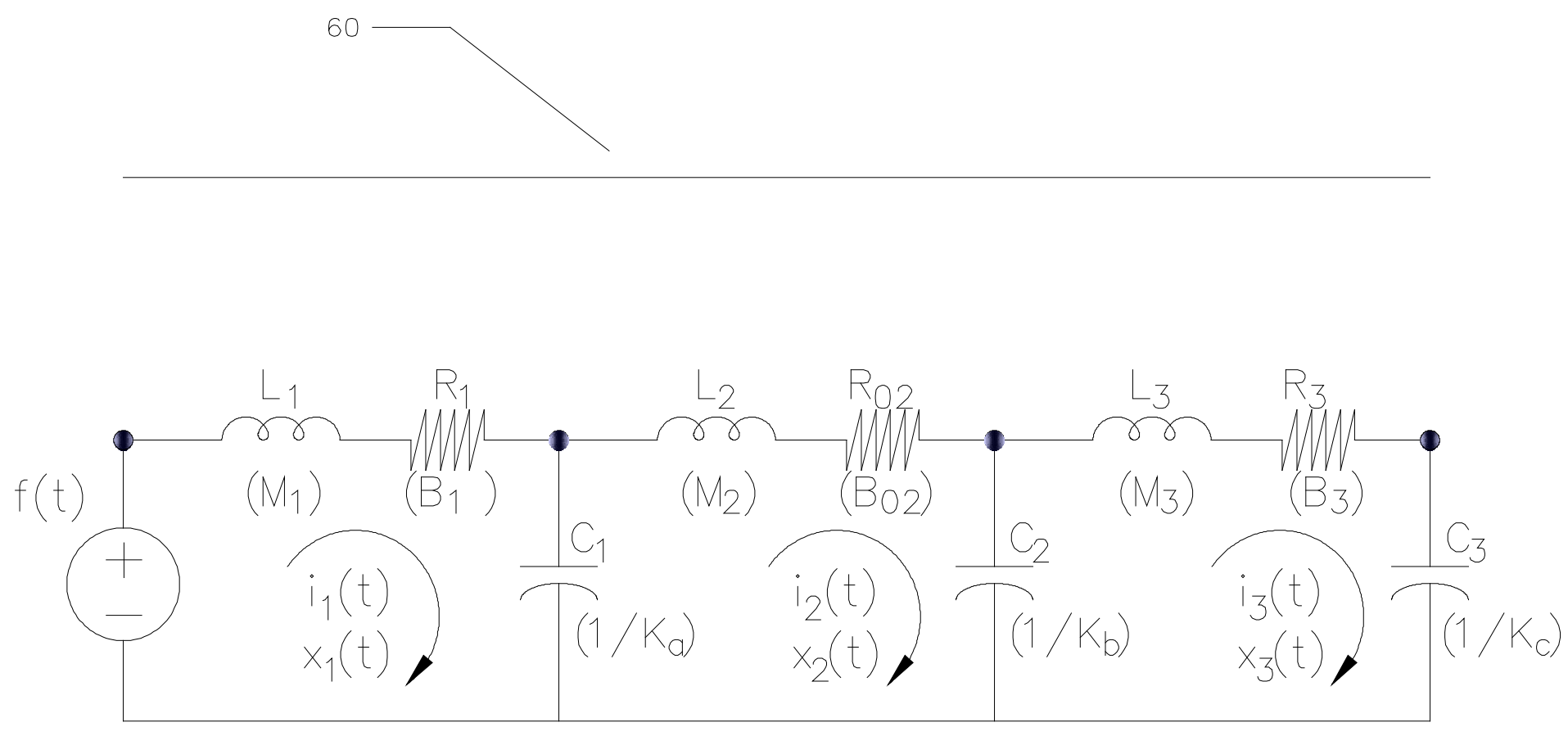

Figure-8 Electrical Modeling

*Legend*

60 - The electrical differential equations modeling the three (3) chasses are obtained from the mechanical system in Figure-7. Krishoff's Voltage Law (KVL) for all three loops of Figure-8 is applied. Second order differenitial equations are set to represent the equipment. The major equations to cover the behavior of the system are (2) through (4).

$$L_1 \frac{di_1}{dt} + Ri_1(t) + \frac{1}{c}\int(i_1 - i_2)dt = f(t) \quad (1)$$

which can be simplified to,

$$L_1 \frac{d^2q_1}{dt^2} + R_1 \frac{dq_1}{dt} + \frac{1}{c_1}(q_1 - q_2) = f(t) \quad (2)$$

$$-\frac{1}{c_1}(q_2 - q_1) + L_2 \frac{d^2q_2}{dt^2} + R_{02} \frac{d_{q2}}{dt} + \frac{1}{c_2}(q_2 - q_1) = 0 \quad (3)$$

$$-\frac{1}{c_2}(q_3 - q_2) + L_3 \frac{d^2q_3}{dt^2} + R_3 \frac{dq_3}{dt} + \frac{1}{c_3}(q_3) = 0$$
(4)

The state variables equation consists of a square 3 X 3 matrix. It can be set up for the combined equations (2), (3) and (4). Equations (2), (3) and (4) are represented in voltage/charge format. All the components of the mechanical system conversion to electrical are depicted in the related figure. However, we're interested only in Unit A2, due to the stability of the other two units A1 and A3.

Therefore, a simple Mass-Spring–Damper system is expressed in (5):

$$M_2\ddot{x} \quad + \quad B_{02}\,\dot{x} \quad + \quad K_b x \quad = \quad f(t) \quad (5)$$

Intertia     Damper     Spring     Applied Force

where,

$M_2$ = Mass (Interia)
$B_{02}$ = Friction–Dissipation
$K_b$ = Stiffness (Spring)
$f$ = Force (N)
$\ddot{x}$ = Acceleration (m/sec$^2$)
$\dot{x}$ = Velocity (m/sec)
$x$ = Displacement (m)

If we let (5) be a homogenous second order DE with constant coefficients, then,

$$M_2\ddot{x} + B_{02}\dot{x} + K_b x = 0$$

This matches the general mathematical linear second order homogeneous model with form:

$$m\frac{d^2x}{dt^2} + c\frac{dx}{dt} + kx = 0$$

The Appendix best provides the parameter estimates for this modeling based on the proposed algorithm.

## Autocorrelation for Damped Oscillation

The autocorrelation function (ACF) often provides insight into the structure of data vectors. The ACF is developed for the second order motion model

$$x(t) = Ce^{-bt} sin(\alpha t + \varphi),$$

where $Ce^{-bt}$ is the time varying amplitude with damping factor $e^{-bt}$ which tends to 0 as t→∞.

## Derivation of ACF

By definition, the autocorrelation function is:

$$R_{xx}(\tau) = \frac{1}{T}\int_0^T x(t)x(t+\tau)dt$$ , where,

$$x(t) = Ce^{-bt}\sin(\alpha\, t + \varphi)$$

$$x(t+\tau) = Ce^{-b(t+\tau)}\sin\left[\alpha\left(t+\tau\right)+\varphi\right]$$

Calculations show by trigonometric identities and change of variable manipulation (and confirmation on Mathematica) of the integral $R_{xx}(\tau)$:

$$R_{xx}(\tau) = \frac{C^2}{T}e^{-b\tau}\left\{\begin{array}{l}\dfrac{\alpha b\sin(\alpha\tau+2\varphi)-b^2\cos(\alpha\tau+2\varphi)+(a^2+b^2)\cos(\alpha\tau)}{4b(\alpha^2+b^2)}\\ -e^{-2bT}\left[\dfrac{\alpha b\sin(2\alpha T+\alpha\tau+2\varphi)-b^2\cos(2\alpha T+2\tau+2\varphi)+(a^2+b^2)\cos(\alpha\tau)}{4b(\alpha^2+b^2)}\right]\end{array}\right\}$$

Normalize $R_{xx}(\tau)$by a constant $K$ for relation by setting τ to 0 such that,

$$K = \frac{1}{R_{xx}(0)}$$

$-1 \le R_{xx}(\tau) \le +1, 0 \le \tau \le N-1$ ($N$ is sample size for discrete observations).

# APPENDIX
# Calculations for Proposed Solution

O'Brien and Johnnie (2011) derived relations for the rapid calculation of the key motion parameters (i.e., amplitude, period, phase, frequency) of noise corrupted SHM via MA filtering. This algorithm can be adapted for damped oscillation signals with modifications unique to damped oscillation modeling. We summarize them here in the Appendix. The system estimates achieve a higher degree of accuracy for $T, \varphi, \alpha$ than the standard "formal" method presented in the main body of the paper. We believe the proposed general method provides an algorithm that is more defensible.

The basis is analytic and computational trigonometry of a right horizontally shifted (time delayed) sine wave, $\sin(\alpha t + \varphi)$, with no vertical shift offset. In essence, we treat this as a periodic function for purposes of modeling. A slightly different notation system is used for these estimates.

Fig. 9 shows the basic graph structure the authors use to calculate the parameters of noise corrupted oscillators. The graph is an idealized solution for noisy data and depends significantly on the noise level. It is used simply to illustrate the calculations in the algorithm. The integer time point markers $t_{k\pi}$ are readily determined numerically from a smoothed graph of empirical data and half-period points $t_{\frac{k}{2}\pi}$ are computed from definitions.

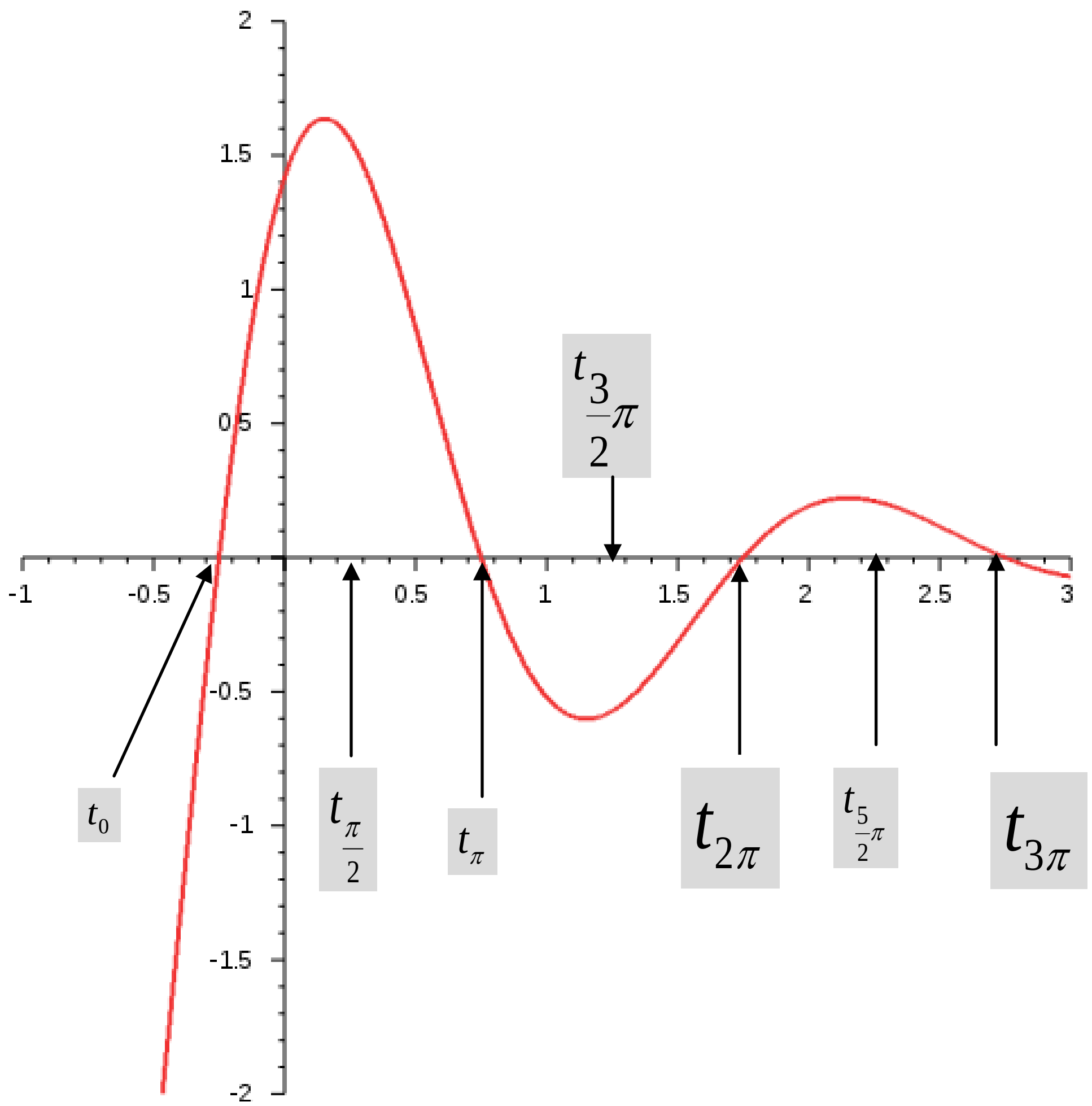

Figure 9. Basic Time-point Relations for Rapid Calculation of Parameters from Hypothetical Smoothed Data Distribution for DE, $x'' + 2x' + (\pi^2 + 1)x = 0$. Only $t_\pi$ and $t_{2\pi}$—or similar half-periods — are read from the graph as inputs; others are calculated. The general graph structure is applicable to any oscillation system with damping. The MA filter zeroes out the first *k*-1 values in actual engineering practice.

The values of the salient time–points of Fig. 9 data are given in the table below—determined by inspection and the following relations:

We first solve a simple simultaneous equation with back substitution for the observed cross-over graph values of $t_\pi = \frac{3}{4}$ & $t_{2\pi} = 1\frac{3}{4}$ (or other half-periods on the graph for sample size *N*) taken from Fig. 9. Experience teaches that these tie-down points can be read from a FIRF MA graph with less numerical error than others. Accurate numerical $t_{\frac{k}{2}\pi}$ values cannot be determined at this point in the algorithm.

This fact is a major shortcoming of standard methods. Averaging more data points across multiple periods will improve accuracy.

One–period, one–half, and one–quarter period can be expressed:

(a) $T = t_{\frac{5}{2}\pi} - t_{\frac{1}{2}\pi}$ (by definition)

(b) $\frac{T}{2} = t_{2\pi} - t_{\pi} = 1$ (input)

(c) $\frac{T}{4} = t_{\pi} - t_{\frac{1}{2}\pi} = \frac{1}{2}$ (by computation)

Other starting points can be used so long as these structural equations are satisfied within the same period. See below for a general set of simultaneous equations.

Thus, from input (b), $2\left(\frac{T}{2}\right) = 2$, and by back substitution we find:

$$t_{\frac{1}{2}\pi} = \frac{3}{4} - \frac{1}{2} = \frac{1}{4} \text{ from (c)}$$

$$t_{\frac{5}{2}\pi} = t_{\frac{1}{2}\pi} + T = 2\frac{1}{4} \text{ from (a),}$$

and, in general, for these data

$$t_{\frac{k}{2}\pi} = \frac{k\pi - 2\varphi}{2\alpha} = t_{\frac{1}{2}\pi} + \frac{k-1}{2} \quad (k = 1,3,5,\ldots)$$

$$t_{k\pi} = \frac{k\pi - \varphi}{\alpha} = t_{\pi} \pm (k-1) \quad (k = 1,2,\ldots)$$

Using the above definitions, a generalized set of equations of (a), (b), (c) can be written as:

- $T = t_{\frac{k}{2}\pi} - t_{\frac{k-4}{2}\pi} \quad (k \text{ is odd}, k \geq 5)$
- $\frac{T}{2} = t_{\frac{k-1}{2}\pi} - t_{\frac{k-3}{2}\pi}$
- $\frac{T}{4} = t_{\frac{k-3}{2}\pi} - t_{\frac{k-4}{2}\pi}$

which reduce, respectively, to $\frac{2\pi}{\alpha}, \frac{\pi}{\alpha}, \frac{\pi}{2\alpha}$. They can be used for computations across multiple points of the MA filter.

Next, we can determine the time delay,

$$\Delta t = T - t_{2\pi} = \frac{1}{4}.$$

As derived in O'Brien/Johnnie and the above, it can be shown that:

$$T = t_{\frac{5}{2}\pi} - t_{\frac{1}{2}\pi} = \frac{5\pi-2\varphi}{2\alpha} - \frac{\pi-2\varphi}{2\alpha} = \frac{4\pi}{2\alpha}, \text{ or}$$

for our data set,

$$\alpha = \pi$$
$$\varphi = \alpha\Delta t = \frac{1}{4}\pi$$

which provides 3 parameters of the model thus far with high accuracy.

**NOTE**: In practice, *T* is also obtained by doubling a half-period, $2\left[t_{k\pi} - t_{(k-1)\pi}\right]$.

**NOTE**: As mentioned earlier in the section on frequency, the envelope function $e(t)$ can be determined by plotting the function $\pm Ce^{-bt}$ with average *t* values,

$$t_{\frac{(2k+1)\pi}{2}} = \frac{t_{\pi k} + t_{(k+1)\pi}}{2} = \frac{(2k+1)\pi - 2\varphi}{2\alpha}$$

which is the same as above with a different domain *k*.

This gives $t_{\frac{(2k+1)\pi}{2}} = k + \frac{1}{4}(k = 0,1,2,\ldots)$ for our data set. These values allow more accurate estimates of *C* & *b*.

Other parameters can be subsequently estimated. These relations for $t_{k\pi}$ and $T, \varphi, \alpha$ provide the following tabulated summary for the general case and the example data set:

| Graph Time Point | | Graph Time Value (Fig. 9 data) | Estimated Graph Function Value $x\left(t_{k\pi}\right)$ |
|---|---|---|---|
| $k$ | $t_{k\pi}$ | $\frac{k\pi - \varphi}{\alpha} = k - \frac{1}{4}$ | $Ce^{-bt_{k\pi}}\sin(\alpha\, t_{k\pi} + \varphi) = Ce^{-b\left(k-\frac{1}{4}\right)}\sin(\pi k)$ |
| 0 | $t_0 = t_{\Delta t}$ | $-\frac{1}{4}$ | 0 (shift factor) |
| $\frac{\varphi}{\pi}$ | $t_{\varphi}$ | 0 (origin) | $\sqrt{2}$ ( $x(0)$ ) |
| $\frac{1}{2}$ | $t_{\frac{1}{2}\pi}$ | $\frac{1}{4}$ | 1.55 to 1.56 to 1.60<br>(Envelope point) |
| 1 | $t_{\pi}$ | $\frac{3}{4}$ | 0 |
| $\frac{3}{2}$ | $t_{\frac{3}{2}\pi}$ | $1\frac{1}{4}$ | –0.60<br>(Envelope point) |
| 2 | $t_{2\pi}$ | $1\frac{3}{4}$ | 0 |
| $\frac{5}{2}$ | $t_{\frac{5}{2}\pi}$ | $2\frac{1}{4}$ | 0.20<br>(Envelope point) |
| 3 | $t_{3\pi}$ | $2\frac{3}{4}$ | 0 |
| **NOTE**: Averaging across periods improves accuracy in the presence of noise. | | | |

The summary parametric relations in the following table—comprising the algorithm—have been adapted for damped oscillation based on SHM modeling in O'Brien & Johnnie (2011). It will be noted that the sequence of calculations is the reverse order of that given in the traditional "formal method". The values for the example data set can be substituted to determine specific calculations based on the structure derived from FIRF filtering. As can be observed $T, \alpha, \varphi$ have a higher accuracy compared to the "formal" method demonstrated in the Specification. *C & b* have about the same or more accuracy and depend on accurate read outs of the FIRF graph.

| Build Sequence ⇓ | Calculation of Parameter Set for $x(t) = Ce^{-bt}\sin(at+\varphi)$ | |
|---|---|---|
| | **FORMULA** | **NOTE** |
| **Period** | $T = 2(t_{2\pi} - t_{\pi})$ or $t_{\frac{5}{2}\pi} - t_{\frac{1}{2}\pi}$ | $\int_T x(t)dt \neq 0$ <br>• <br>• Other starting points possible |
| **Phase angle** | $\varphi = \frac{2\pi\,\Delta t}{T}$ <br>where $\Delta t = T - t_{2\pi}$ and $t_{k\pi} = \frac{k\pi - \varphi}{\alpha}$ <br>$(\Delta t \neq 0)$ | • General unifying formula <br>• $T = \frac{2\pi}{\alpha}$ <br>• $\lvert\Delta t\rvert$ is phase displacement |
| **Frequency** | $\alpha = \frac{\varphi}{\Delta t}$ or $\frac{2\pi}{T}$ $(\Delta t \neq 0)$ | $\frac{\varphi}{\alpha}$ <br>is the time delay |
| **Damping Coefficient** | $b = \frac{1}{t_2 - t_1}\ln\left[\frac{Ce^{-bt_1}}{Ce^{-bt_2}}\right]$ $t_1 = t_{\frac{k}{2}\pi};\, t_2 = t_{\frac{k+4}{2}\pi}$ ($k$ odd) | Modified prior art |
| **Constant** | $C = \frac{1}{2}\left[\frac{Ce^{-bt_1}}{e^{-bt_1}} + \frac{Ce^{-bt_2}}{e^{-bt_2}}\right]$ <br>$t_1 = t_{\frac{k}{2}\pi};\, t_2 = t_{\frac{k+4}{2}\pi}$ ($k$ odd) | Modified prior art |
| **NOTE**: Overall we obtain an average error rate of about 1-2%, which is less than the traditional method. Averaging across multiple points of the periods improves estimation stability especially in the presence of noise. Averaging formulas are presented below. | | |

**NOTE:** Phase can be calculated anywhere on the graph depending upon the measurements available (or missing) by the following derived relations:

$$\varphi_{rad} = \begin{cases} \frac{T - t_{2\pi}}{t_{k\pi} - t_{2\pi}}(k\pi - 2\pi), & (k \neq 2) \\ \frac{\pi\,\Delta t}{t_{k\pi} - t_{(k-1)\pi}}(k > 0) \end{cases}$$

$$\varphi^{\circ} = \frac{\varphi_{rad}}{\pi}180$$

## Averaging Formulas for Damped Harmonic Oscillation Parameters

| AVERAGING FORMULA | NOTE |
|---|---|
| $\overline{T} = \frac{2}{K-1}\sum_{k=2}^{K}\left(t_{k\pi} - t_{(k-1)\pi}\right)$<br>$= \frac{1}{K-1}\left\{\begin{matrix} 2(t_{2\pi}-t_{\pi})+2(t_{3\pi}-t_{2\pi})+2(t_{4\pi}-t_{3\pi}) \\ +\ldots+2(t_{K\pi}-t_{(K-1)\pi}) \end{matrix}\right\}$ | $\overline{T} \to T = \frac{2\pi}{\alpha}$, assuming constants $\alpha$ & $\varphi$. |
| $\overline{\Delta}t = \frac{1}{K}\sum_{k=1}^{K}\left(\frac{k}{2}T - t_{k\pi}\right)$<br>$= \frac{1}{K}\left\{\left(\frac{1}{2}T - t_{\pi}\right)+\left(T - t_{2\pi}\right)+\left(\frac{3}{2}T - t_{3\pi}\right)+\ldots+\left(\frac{K}{2}T - t_{K\pi}\right)\right\}$ | $\overline{\Delta}t \to \Delta t = \frac{\varphi}{\alpha}$, assuming constants $\alpha$ & $\varphi$. |
| $\overline{b} = \frac{1}{(K+1)}\sum_{k=0}^{K}\frac{1}{\frac{T_k}{2}}\ln\left[\left\lvert\frac{Ce^{-bt_{(2k+1)\frac{\pi}{2}}}}{Ce^{-bt_{(2k+3)\frac{\pi}{2}}}}\right\rvert\right]$<br>$= \frac{1}{(K+1)}\left\{\begin{matrix} \frac{1}{\frac{T_0}{2}}\ln\left[\left\lvert\frac{Ce^{-bt_{\frac{\pi}{2}}}}{Ce^{-bt_{\frac{3\pi}{2}}}}\right\rvert\right]+\frac{1}{\frac{T_1}{2}}\ln\left[\left\lvert\frac{Ce^{-bt_{\frac{3\pi}{2}}}}{Ce^{-bt_{\frac{5\pi}{2}}}}\right\rvert\right]+\frac{1}{\frac{T_2}{2}}\ln\left[\left\lvert\frac{Ce^{-bt_{\frac{5\pi}{2}}}}{Ce^{-bt_{\frac{7\pi}{2}}}}\right\rvert\right] \\ +\ldots+\frac{1}{\frac{T_K}{2}}\ln\left[\left\lvert\frac{Ce^{-bt_{(2K+1)\frac{\pi}{2}}}}{Ce^{-bt_{(2K+3)\frac{\pi}{2}}}}\right\rvert\right] \end{matrix}\right\}$ | • $k$ averages integer odd values over half-periods $T/2$:<br>$\left(\frac{1}{2}\pi,\frac{3}{2}\pi\right),\left(\frac{3}{2}\pi,\frac{5}{2}\pi\right),\left(\frac{5}{2}\pi,\frac{7}{2}\pi\right)$....<br>Absolute values since pairings of num. & denom. are unlike signs.<br>• $t_{(2k+3)\frac{\pi}{2}} - t_{(2k+1)\frac{\pi}{2}} = \frac{T_k}{2}$ |
| $\overline{C} = \frac{1}{2(K+1)}\sum_{k=0}^{K}\left[\left\lvert\frac{Ce^{-bt_{(2k+1)\frac{\pi}{2}}}}{e^{-bt_{(2k+1)\frac{\pi}{2}}}}\right\rvert+\left\lvert\frac{Ce^{-bt_{(2k+3)\frac{\pi}{2}}}}{e^{-bt_{(2k+3)\frac{\pi}{2}}}}\right\rvert\right]$<br>$= \frac{1}{(K+1)}\left\{\begin{matrix} \frac{1}{2}\left[\left\lvert\frac{Ce^{-bt_{\frac{\pi}{2}}}}{e^{-bt_{\frac{\pi}{2}}}}\right\rvert+\left\lvert\frac{Ce^{-bt_{\frac{3\pi}{2}}}}{e^{-bt_{\frac{3\pi}{2}}}}\right\rvert\right]+\frac{1}{2}\left[\left\lvert\frac{Ce^{-bt_{\frac{3\pi}{2}}}}{e^{-bt_{\frac{3\pi}{2}}}}\right\rvert+\left\lvert\frac{Ce^{-bt_{\frac{5\pi}{2}}}}{e^{-bt_{\frac{5\pi}{2}}}}\right\rvert\right] \\ +\ldots+\frac{1}{2}\left[\left\lvert\frac{Ce^{-bt_{(2K+1)\frac{\pi}{2}}}}{e^{-bt_{(2K+1)\frac{\pi}{2}}}}\right\rvert+\left\lvert\frac{Ce^{-bt_{(2K+3)\frac{\pi}{2}}}}{e^{-bt_{(2K+3)\frac{\pi}{2}}}}\right\rvert\right] \end{matrix}\right\}$ | • $k$ averages integer odd values over half-periods,<br>$\left(\frac{1}{2}\pi,\frac{3}{2}\pi\right),\left(\frac{3}{2}\pi,\frac{5}{2}\pi\right),\left(\frac{5}{2}\pi,\frac{7}{2}\pi\right)$....<br>where summands are absolute values since $e(t)$ & $-e(t)$ are averaged in each loop such as $t_{\frac{1}{2}\pi}$ & $t_{\frac{3}{2}\pi}$ in first term of summation, $k$ = 0. |

**NOTE**: these formulas follow from Appendix formulations.

**NOTE**: the formulas for $\overline{T}$ and $\overline{\Delta}t$ allow best estimates for phase angle $\varphi$ and angular frequency $\alpha$.

**NOTE**: proofs for the formulas derived by evaluating the summations using the definition of $t_{k\pi}$. For example, the expansion for period $T$ shows:

$$\overline{T} = \frac{2}{K-1}\sum_{k=2}^{K}\left(t_{k\pi} - t_{(k-1)\pi}\right)$$

$$= \frac{2}{K-1}\sum_{k=2}^{K}\left(\frac{k\pi-\varphi}{\alpha} - \frac{(k-1)\pi-\varphi}{\alpha}\right)$$

$$= \frac{2}{K-1}\sum_{k=2}^{K}\left(\frac{k\pi}{\alpha} - \frac{(k-1)\pi}{\alpha}\right)$$

$$= \frac{2\pi}{\alpha}\frac{1}{(K-1)}\sum_{k=2}^{K}(1) = \frac{2\pi}{\alpha(K-1)}(K-1)$$

$$= \frac{2\pi}{\alpha} = T.$$

Thus $\overline{T} \to T$ if $\alpha$ & $\varphi$ are assumed constant.

Other proofs follow in a similar manner.

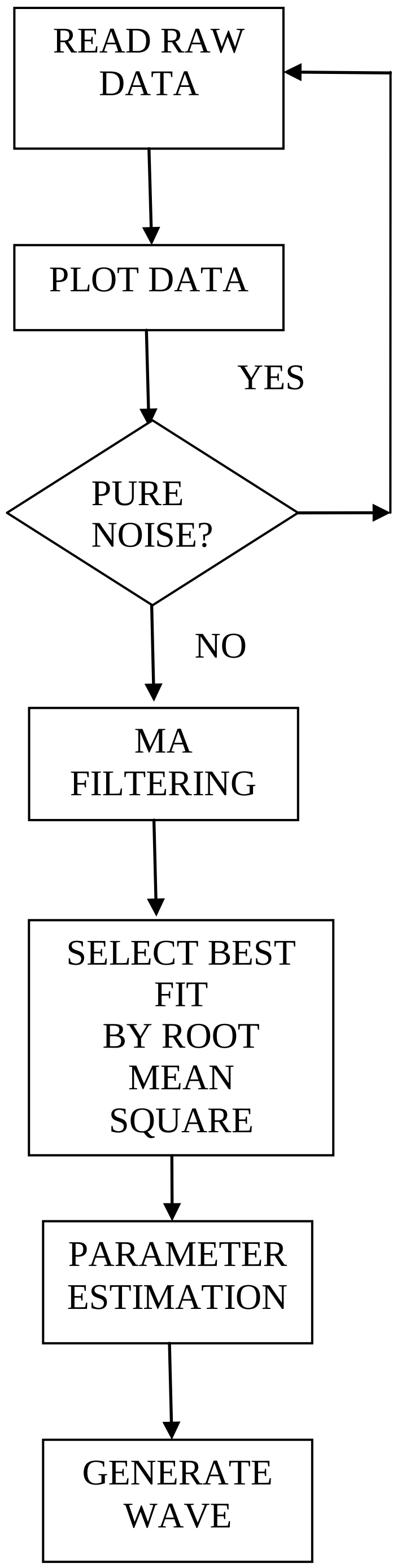

**METHOD FLOW CHART**